\documentclass[12pt,superscriptaddress,nofootinbib]{revtex4}

\usepackage{amssymb,amsmath,amsfonts,amsthm,graphicx,psfrag}
\usepackage{hyperref}
\usepackage{slashed,braket}

\newcommand{\bbraket}[1]{\braket{\braket{#1}}}
\newcommand{\tr}{\text{tr}}
\newcommand{\Tr}{\text{Tr}}
\newcommand{\beq}{\begin{eqnarray}}
\newcommand{\eeq}{\end{eqnarray}}
\newcommand{\be}{\begin{equation}}
\newcommand{\ee}{\end{equation}}

\newcommand{{\cD}}{\mathcal{D}}
\newcommand{{\const}}{\text{~const~}}
\newcommand{{\MeV}}{\text{~MeV~}}
\newcommand{\vex}{\mbox{\boldmath${\rm x}$}}

\newcommand{\ver}{\mbox{\boldmath${\rm r}$}}
\newcommand{\vep}{\mbox{\boldmath${\rm p}$}}
\newcommand{\veu}{\mbox{\boldmath${\rm u}$}}
\newcommand{\lan}{\langle}
\newcommand{\ran}{\rangle}

\begin{document}

\title{Quark condensate from confinement in QCD}

\author{R.A.Abramchuk}
\email{abramchuk@phystech.edu}
\affiliation{Institute for Theoretical and Experimental Physics of NRC ``Kurchatov Institute'', B. Cheremushkinskaya 25, Moscow, 117259, Russia}
\affiliation{Moscow Institute of Physics and Technology, 9, Institutskii per., Dolgoprudny, Moscow Region, 141700, Russia}

\author{Yu.A.Simonov}
\email{simonov@itep.ru}
\affiliation{Institute for Theoretical and Experimental Physics of NRC ``Kurchatov Institute'', B. Cheremushkinskaya 25, Moscow, 117259, Russia}

\begin{abstract}
    The scalar confinement in QCD is shown to produce the nonzero quark condensate for any current quark mass. 
    Mechanisms for the Chiral Symmetry breaking and for the nonzero quark condensates are revealed.
    For the light and strange flavors the condensates are essentially chiral (proportional to (string tension)$^{3/2}$).
    The expressions for the quark condensates are obtained as a sum over non-chiral meson states,
        and, using the operator product expansion, as an explicit expression
            that involves the confining string tension and the current quark mass.
    The numerical results are fairly close to the lattice data.
\end{abstract}
\maketitle

\section{Introduction} \label{SectIntro}

%\red{THE PRINCIPAL AIM}
The chiral symmetry breaking (CSB) is the fundamental property of the QCD dynamics \cite{1,2,3} 
(see the reviews \cite{4} and the recent summaries of the present knowledge on this subject \cite{5,6}).
The primary parameter of CSB is the quark condensate $ \Delta_i = | \lan \bar q_i q_i\ran |$,
    which is nonzero in the chiral broken phase.
Lack of explanation for CSB dynamical origins is a major problem of QCD, as well as evaluation of the parameter.
Various theoretical models were suggested to implement chiral degrees of freedom (d.o.f.) and nonzero $\Delta_i$, see e.g. \cite{7}.

A natural indirect way to obtain $\Delta_i$ from lattice simulations is the use of the Gell-Mann-Oakes-Renner (GMOR) relations \cite{14}
(GMOR relates $\Delta_i$ with meson masses, decay constants and current quark masses \cite{15}).
Another approach is computation of $\Delta_i$ on lattice via operator product expansion (OPE).
The approach provides a good accuracy \cite{16,17}.

The conventional analytic approach to the chiral condensate evaluation is based on the Banks-Casher relation \cite{10},
    which expresses \(\Delta\) via the vacuum Dirac operator spectral density $\rho(E\to +0)$ near the zero energy level.
The instanton model \cite{8,9} relies on the property of an isolated instanton to provide a zero-mode for the Dirac operator.
So the instanton gas medium may generate the nonzero Dirac operator spectral density near the zero energy level.
The lattice computations seemingly can be adjusted to this approach \cite{11},
    but the results disagree with the values computed in other ways \cite{11,12,13}.

In this paper, we utilize the Chiral Confining theory (CCTh) \cite{18,19,20,21,22}.
The confining vacuum in CCTh emerges from the background stochastic gluon field.
While the multi-instanton vacuum is a fine-tuned collective model,
   the confining vacuum in CCTh is a self-consistent approximate solution of gluodynamics and QCD \cite{24}.

The `fundamental' d.o.f. in CCTh physical picture is an individual quark
    that moves in the confining vacuum.
A gauge invariant definition of a Dirac operator eigenmode requires construction of a white object 
   that is the quark propagating along a closed trajectory in the 4D Euclidean space-time. 
The associated Wilson loop obeys the area law,
    which produces non-perturbative quark effective mass.
The effective mass is nonzero even for zero current quark mass,
    which dynamically brakes the chiral symmetry and spawns the quark condensate.
Therefore, the Dirac operator low-energy eigenmodes 
    , which are crucial for the Banks-Casher approach, 
    are irrelevant. 
The main problem for us is an accurate calculation of the quark and meson Green's functions,
    which we address with the Fock-Feynman-Schwinger Method \cite{23}.

Scalar character of the confinement is also a necessary condition for the bound states stability \cite{24'}.
As we show in the present paper, the quark condensate is nonzero and the chiral symmetry is broken in the QCD confining vacuum,
    or the nonzero confining string tension yields nonzero quark condensates
(indeed, in lattice simulations, the pseudo-critical temperatures for the chiral and deconfinement transitions coincide within error \cite{25}).

The effective action for CCTh is the Chiral Confining Lagrangian CCL \cite{19,20,21}
\be
    L_{eff} (\hat M, \varphi)=- N_c \Tr \log [ i\slashed \partial + \hat m + \hat M \hat U], \label{1}
\ee
where $\hat U(\varphi)= \exp(i\gamma_5 \Phi(\varphi))$,
    $\Phi$ is the standard chiral matrix 
         that is a linear combination of meson fields $\varphi(x)$
         , and, finally, the operator $M(r) = \sigma r$ represents the confining interaction. 
The simple decomposition $\hat M \hat U= \hat M + \hat M (\hat U - 1)$ separates the standard $q-\bar q$ confining interaction in the first term,
    and transitions from $q-\bar q$ to one or more chiral mesons in the second term.

CCL yields all known chiral relations, including the GMOR relations \cite{14'}.
Moreover, CCL allows to calculate the decay constants $f_i$  \cite{26} as the sum over excited hadron states \cite{20,21}
(the results of \cite{26} for $f_\pi, f_k$ are in a good agreement with experimental data).
CCL also allows to obtain the chiral perturbation theory corrections in $p^4$ order \cite{27} and in higher orders.
Finally, CCL correctly describes the dependence of $\Delta_i,~ f_i$ and chiral hadron masses on the magnetic field \cite{28}, while the standard chiral theory strongly disagrees with lattice data.

In \cite{21} the quark condensate $\Delta_i$ was calculated as a sum over all non-chiral pseudo-scalar (PS) bound states,
    which are produced by the linear confining, color Coulomb and spin-dependent interactions.
However, a convergence analysis for the sum was missing. 
In this paper we suggest a series truncation procedure that provides a satisfactory numerical result,
    as well as a well-defined regularization procedure.

Then we present another calculation of $\Delta_i$ using the Operator Product Expansion (OPE).
With this method we treat the non-perturbative confining interaction, and perform the standard renormalization \cite{29}.
Within this approach, the condensates \(\Delta_i\) are expressed via the confining string tension $\sigma$ and quark masses $m_q$.

The plan of the paper is as follows. 
Section \ref{SectConf} contains a definition of the quark propagator in the confining vacuum field,
    which is a set of integral equations for the averaged propagator and the effective quark mass operator. 
In Section \ref{SectCCL} we derive the expression in form of a sum over PS states for $\Delta_i$ from CCL following the original derivation of \cite{21}.
In Section \ref{SectFFSR} we calculate the series terms in the linear confinement approximation.
In Section \ref{SectSpectrSum} we extract numerical values from the sum and analyze the sum structure.
A more general and explicit derivation of the quark condensate, its dependence on quark mass is done with the non-perturbative OPE in Section \ref{SectOPE}.
Section \ref{SectConc} contains discussion and conclusions.

\section{From scalar confinement to spontaneous violation of chiral symmetry} \label{SectConf}

In this section we show following \cite{18} 
    that the vacuum averaging of the confining vacuum fields leads, in the lowest cumulant order, to the set of two connected equations for the averaged quark Green's function and the vacuum averaged effective quark mass operator,
        which we use to calculate the quark condensate.

We apply cumulant expansion directly to QCD Lagrangian
\begin{gather}
    \int\cD A \exp\int d^4x \left(\psi^\dag\slashed{A}\psi + \frac12\tr F_{\mu\nu}^2\right) =
    \braket{\exp\int d^4x \left(\psi^\dag\slashed{A}\psi\right)}_A = \\
    = \exp\left(\int d^4x \psi^\dag(x)\gamma_\mu\bbraket{A_\mu(x)}\psi(x)\right. \\
                \left. + \frac12\int d^4xd^4y \psi_a^\dag(x)\gamma_\mu\psi_b(x) \psi_c^\dag(y)\gamma_\mu\psi_d(y)
                \bbraket{A_\mu^{ab}(x)A_\mu^{cd}(y)}+\ldots
           \right),
\label{2}
\end{gather}
where the Latin indices refer to the fundamental gauge group.
Now using the Fock-Schwinger gauge
\be
    A_\mu(x) = \int_0^1du~ux_\nu F_{\nu\mu}(ux)
\label{3} \ee
we get
\be
    \bbraket{A_\mu^{ab}(x)A_\nu^{cd}(y)} = \frac{\delta_{bc}\delta_{ad}}{N_c} \int_0^1du~ux_\lambda \int_0^1dv~vy_\rho
    \bbraket{\tr F_{\lambda\mu}(ux)F_{\rho\nu}(vy)}.
\label{4} \ee
The latter correlator may be expressed via the standard correlators $D(z),D_1(z)$ \cite{30}. 
We keep for simplicity only the confining component $D(z)$ (since the non-confining interaction does not generate quark condensates)
\be
    \braket{\tr F_{i4}(x)F_{j4}(y)} = -\delta_{ij}D(x-y).
\label{5}\ee

As a result, we obtain the effective quark action in the confining vacuum
\begin{align}
    \mathcal{L}_\text{eff}(\psi^\dag, \psi) = &\int d^4x\psi^\dag(x)(-i\slashed{\partial}-im)\psi(x) +\\
    &+ \frac{1}{2N_c}\int d^4xd^4y~\psi^\dag_a(x)\gamma_4\psi_b(x)~\psi_b^\dag(y)\gamma_4\psi_a(y)~J(x,y),
\end{align}
\label{6} 
where
\be
    J(x,y) = \int_0^1du~x_i\int_0^1dv~y^i D(ux-vy).
\label{7}\ee

The quark propagator \(S(x,z)\) and the effective quark mass $M(x,y)$ for this effective action satisfies the set of equations
\begin{gather}
    iM(x,z) = J(x,z)\gamma_4 S(x,z)\gamma_4,\\
    (\slashed{\partial}+m)S(x,y) + \int d^4z M(x,z)S(z,y) = \delta(x-y).
\label{8}
\end{gather}
In what follows we use the local approximation of the effective mass operator
\be
    M(x,z) \approx ~\sigma\lambda\delta(x-z)
\label{9}\ee
that yields an  approximation for the effective quark propagator
\be
    S^{-1} = \slashed{\partial}+m+M,\quad M=\sigma\lambda
\label{10}\ee
where $\lambda$ is the average length of the confining string.

\section{Quark condensate from the Chiral Confining Lagrangian} \label{SectCCL}

Here we follow the reasoning of \cite{21}, and write the chiral condensate as
 \be\Delta_i =  N_c |tr S_{xx}|;~~ S_{xx}=\left( \frac{1}{m+ \hat M + \slashed \partial}\right)_{xx}.\label{11}\ee
At this point, two roles  of the operator $\hat M$ in (\ref{11}) should be distinguished. 
It can enter as the \(\bar q-q\to\) mesons vertex from \(\hat M(\hat U-1)\), 
    in which  case $\hat M = M(\lambda) = \sigma \lambda =0.15$ GeV \cite{18,19,20,21,22}, 
    while on the trajectory of the quark $\hat M$ provides the scalar confinement, $M(r) =\sigma  r$, 
         where  $r$ is the distance to the center of the closed trajectory, 
             which is equal to $\lambda$ on average.

Dividing  and  multiplying (\ref{11}) by $(m+ M(r) - \slashed \partial)$, we obtain
\begin{align}
    \tr S(x,x) &= \tr \left( \frac{1}{(m+M+\slashed \partial)}\cdot  (m+M-\slashed \partial)\frac{1}{(m+M-\slashed \partial)}\right)\\
    &= \int tr (\gamma_5  S(x,y) \gamma_5 (m+M) S(y,x)) d^4 y \\
    &=\braket{m+M} \int \tr (\gamma_5  S(x,y) \gamma_5  S(y,x)) d^4 y \label{12'}\\
    &=\braket{m+ M} \int G (x,y) d^4y,\label{12}
\end{align}
where we recognize $G(x,y)$ as the $q\bar q$ Green's function in the PS channel.
Here $\braket{m+ M}= m+ \braket{M}=m+ \sigma\lambda$. 
In the case $m_q=0$, $\lambda$ is defined by the only dimensionful parameter $\sigma$ as $\lambda\sim\sigma^{-1/2}$. 
As for the growing $m_q$, $\lambda$ is decreasing (see section \ref{SectOPE}).

Expansion of $G^{(\mu)} (x,y)$ in the  complete set of the PS $q\bar q$ eigenfunctions reads \cite{21}
\be G^{(\mu)} (k) = - 2 \sum^\infty_{n=0}  \frac{c_n c_n^{(M)}}{k^2+ m^2_n},\label{13}\ee
where $c_n = \sqrt{ \frac{m_n}{2}} \varphi_n (0)$, and $c^{(M)}_n=(m+M(\lambda)) \sqrt{ \frac{m_n}{2}} \varphi_n (0)$. 
As a result, we obtain the condensate as a spectral sum over PS meson states
\be
\Delta_i= N_c (m+ M(\lambda)) \sum^\infty_{n=0} \frac{|\varphi_n (0)|^2}{m_n}. \label{14}\ee
As we shall see in the next section, this result can be obtained by an independent method.

\section{Chiral condensate from the quark  propagator in the  confining vacuum} \label{SectFFSR}

Let us pursue the direct calculation of \eqref{12'} in the approximation of linear confinement between quarks.
Thus,
\be \Delta_i = N_c (m_q + M(\lambda)) \left( tr \frac{1}{(m_q+M)^2-\partial^2}\right)_{xx} = N_c (m_q+M(\lambda)) f(\sigma, m_q)\label{16}\ee
describes quark propagation along a closed orbit. 
The closed `scalar propagator' \(f(\sigma,m_q)\) in the Feynman-Fock-Schwinger representation reads
\be f(\sigma, m_q) = \int^\infty_0 ds G(s), ~~ G(s)= \int (\cD^4z)_{xx} e^{-K-m^2_q s} \lan \tr W (C_{xx})\ran\label{17}\ee
where \be ( \cD^4 z)_{xx} = \int \frac{d^4p}{(2\pi)^4} \prod^N_{n=1} \frac{d^4\Delta z(n)}{(4\pi \varepsilon)^2} e^{ip \sum\Delta z(n)}, ~~ N\varepsilon =s\label{18}\ee
\be K=\frac14 \int^s_0 \left( \frac{dz_\mu (\tau)}{d\tau} \right)^2 d\tau, ~~ W(C_{xx}) = \exp (ig \int_{C_{xx}} A_\mu  (z) d z_\mu).\label{19}\ee

To do  the integrals in (\ref{18}) we split the contour $C_{xx}$ in  two parts: $(xu) $ and $(u x)$, where $u_\mu = z_\mu (\tau = s/2)$
\begin{gather}
    G(s) = \int (\cD^4z) _{xu} d^4u (\cD^4z')_{ux} e^{- K_1 - K_2-V}, \label{3.6}\\
    K_1=\frac14 \int^{s/2}_0 d\tau \left( \frac{dz_\mu (\tau)}{d\tau} \right)^2 , 
    ~~ K_2=\frac14 \int^{s }_{s/2} d\tau' \left( \frac{dz'_\mu (\tau')}{d\tau'} \right)^2,\\
    V=\int^{s/2}_0\sigma |{\bf z} (\tau)-{\bf z'}(s-\tau)|dz_4(\tau),\quad 
\end{gather}
and turn to the Hamiltonian formalism --- separate time and spatial coordinates 
$(\cD^4z)_{xu} = (\cD^3 z)_{xu} (\cD z_4)_{x_4u_4}$, 
though the space is O(4) invariant.

In our approximation, we neglect back-propagation of the quarks in time.
Then we immediately calculate the temporal parts of the path integrals.
With the $\omega$-representation $ d\tau = \frac{dt}{\omega}$, $\frac{s}{2} = \frac{u_4}{2\omega}, ~~ ds = \frac{u_4 d\omega}{\omega^2}$, 
path-integration over time is trivial
\begin{align}  
    I_4^{(i)} &= \int (\cD z_4)_{x_4u_4} \exp\left(-\frac14 \int^{s/2}_0 d\tau \left( \frac{dz_4}{d\tau}\right)^2\right) \\
    &= \int \prod \frac{\Delta z_4 (n)}{\sqrt{4 \pi\varepsilon}} \exp\left(-\frac14 \frac{(\Delta z_4)^2}{\varepsilon} \right)\frac{d p_4}{2\pi} e^{ip_4  \sum(u_4-x_4)}
    = \sqrt{\frac{\omega}{2\pi u_4}}. 
\end{align}

As a result, the problem is reduced to quantum mechanics
\begin{align}
    f(\sigma,m_q) &= \int d^4u \frac{u_4 d\omega}{\omega^2} \frac{\omega}{2\pi u_4} (\cD^3z)_{xu} (\cD^3z')_{ux} e^{-K_1^{(3)} - K_2^{(3)} -V  - \frac{m^2_qu_4}{\omega}}\\
    &=\int \frac{d^4u}{2\pi} \int^\infty_0 \frac{d\omega}{\omega} \lan \vex,\vex | e^{-\int^{u_4}_0 H dt}|\veu,\veu\ran,
    \quad H = -h_1-h_2-V- \frac{m_q^2 u_4}{\omega};\\
    &= \int^\infty_0 \frac{d u_4}{2\pi} \int^\infty_0 \frac{d\omega}{\omega} \lan 0 | e^{- H_{CM} u_4  }|0\ran \label{21},
    \quad H_{CM} = \frac{\vep^2+m^2_q+\omega^2}{\omega} + V(\ver),
\end{align}
where we expressed the path integrals as an evolution operator matrix element,
    and integrated out the center of mass motion, 
        so reduced the problem to the one-particle motion.

With the spectra of the \(q\bar q\) Hamiltonian for the confining interaction $V(\ver)=\sigma|\ver|$
\be H_{CM} \varphi_n (\ver) = M_n (\omega) \varphi_n (\ver),\label{24}\ee
$f(\sigma, m_q)$ reads
\be f(\sigma, m_q) = \int^\infty_0 \frac{du_4}{2\pi} \int^\infty_0 \frac{d\omega}{\omega} \sum^\infty_{n=0} |\varphi^2_n (0)| e^{-M_n (\omega) u_4}.\label{25}\ee

We calculate the integral over \(\omega\) in the saddle-point approximation
$M_n(\omega_0) \approx M_n (\omega_0) + \frac{(\omega-\omega_0)^2}{2} M_n^{''} (\omega_0)$ around $\omega= \omega_0,~M'_n(\omega_0) =0$
\begin{align} 
    f(\sigma, m_q) &= \int^\infty_0 \frac{du_4}{2\pi}  \sum_n  |\varphi^2_n (0)| e^{-M_n (\omega_0) u_4} \sqrt{\frac{2\pi}{ M_n^{''} (\omega_0) u_4 \omega^2_0}}  \label{26}\\
    &=  \sum^\infty_{n=0} \frac{|\varphi^2_n (0)|}{\sqrt{ 2M_n  M_n^{''}\omega^2_0 }}.\label{27}
\end{align}

Finally, with the approximation $M_n^{''} \omega^2_0 \approx \frac12 M_n$, we obtain the result of the previous section
\be f(\sigma, m_q) =  \sum^\infty_{n=0} \frac{|\varphi^2_n (0)|}{ M_n }.\label{28}\ee

\section{Chiral condensate value from the spectral sum} \label{SectSpectrSum}

In this section we extract a numerical result from the sum over the non-chiral PS states $n$ 
\be \Delta_i = N_c (m_i + M(\lambda)) f(\sigma, m_i), ~~f(\sigma, m_i) =  \sum^\infty_{n=0} \frac{|\varphi_n (0)|^2}{ M_n }.\label{29}\ee
The sum is divergent, since a propagator has a pole at zero distance, and we need to eliminate this `trivial' divergence.

Let us begin with the massless quarks limit,
    where terms $O(\alpha_s)$ may be neglected as the first approximation. 
So, in this simplest approximation
\be f(\sigma, m_i \to 0 , \alpha_s \to 0 ) = c_i\sigma\label{30}\ee
only the non-perturbative part retains.
However, the dimensionless constant $c_i$ may diverge.

As the next approximation, we add color-Coulomb and spin-spin interactions to the linear confining potential. 
Then $\varphi_n(0)$ reads
\be | \varphi_n(0)|^2 = \frac{\omega_n \left(\sigma + \frac43 \alpha_s \lan \frac{1}{r^2}\ran \right)}{4\pi}\label{31}.\ee
The total mass $m_n \approx 2 \omega_n$, where $\omega_n$ is  the effective energy. 
$m^2_n =O(n)$ for large $n$, thus $ \frac{|\varphi_n (0)|^2}{ m_n }  = O(n^0)$, and the sum over $n$ diverges.

Nevertheless, we calculate the first  three terms $n=0,1,2$ following \cite{21}, 
    and treat separately the problem of convergence for the rest terms.

We use the estimates of masses and $\varphi^2_n (0)$ from \cite{21}
 \be m_0 =0.4 ~{\rm GeV},~~ m_1 =1.35 ~{\rm GeV},~~ m_2 =1.85 ~{\rm GeV}, \label{32}\ee
 \be \varphi^2_0 (0) = \frac{0.109}{4\pi} ~{\rm GeV}^3, ~~ \varphi^2_1 (0) = \frac{0.097}{4\pi} ~{\rm GeV}^3, ~~ \varphi^2_2 (0) = \frac{0.115}{4\pi} ~{\rm GeV}^3, \label{33}\ee
which omit the chiral effects, since the spectral sum is over the complete set of non-chiral states \eqref{24}, \eqref{21}.
So, $m_0$ is the spin-spin  companion of $\rho$-meson, and $m_1$ and $m_2$ are not far from the real excited pion states (1.3 GeV and 1.83 GeV in experiment).
The truncated to the three terms series yields the result
 \be f^{(3)} (\sigma, 0)= 0.032 ~{\rm GeV^2} = 0.18 \sigma, ~~ c_0^{(3)}=0.18 \label{34}\ee
 \be \Delta_i^{(3)} = N_c \sigma \lambda f^{(3)} (\sigma, 0) = (272 {\rm ~MeV})^3;~~ \sigma \lambda =0.15 ~{\rm  GeV}.\label{35}\ee

We now turn to the higher terms of (\ref{29}). 
The closed propagator \(f(\sigma,m_q)\) is divergent because of the zero distance pole.
A meaningful result for the chiral condensate can be obtained by subtraction of the regularized pole value of the free propagator.
To do that, let us consider the propagator at a small (temporal) distance \(t_0\)
\be
    f(\sigma, 0, t_0) = \int_{t_0}^{+\infty}dt\sum_n e^{-M_nt}|\varphi_n^2(0)|.
\ee
With color-Coulomb and spin-spin interactions neglected, we utilize the string spectrum 
    \(M_n^2=4\pi\sigma(n+3/4)\), \(|\varphi_n^2(0)| = \frac{\xi\sigma M_n}{4\pi}\) as a model for the spectrum of \eqref{29},
        and approximate the \(n\) summation with the \(dM\) integration
\be
    f(\sigma, 0, t_0) \approx \frac{\xi}{8\pi^2}\int_{M_0}^{+\infty}M^2dM\int_{t_0}^{+\infty}e^{-Mt}dt = \frac{\xi}{8\pi^2}(t_0^{-2}-M_0^2) +O(t_0).
\ee
Comparison with the small-distance free particle propagator behavior \(G(t)\sim(4\pi^2t^2)^{-1}\)
    fixes the parameter \(\frac{\xi}{8\pi^2}=\frac{1}{4\pi^2}\).
Subtraction of the `temporal divergence' leads us to a numerical result, 
    which happens to be fairly close to the lattice data
\be
    f(\sigma,0) = -\frac{M_0^2}{4\pi^2} = -\frac{3\sigma}{4\pi},
\ee
\be
    \Delta_l = \frac{3N_c\sigma^2\lambda}{4\pi} = (268 \text{ MeV})^3 \label{29'}.
\ee

\section{Quark condensate from the OPE} \label{SectOPE}

We consider the quark Green's function in the Euclidean space-time in the confining field following the
methods introduced in \cite{29}
\be
\left(\slashed\partial + m_q + \sigma |{\bf x}|\right) S(x,y)= \delta ^{(4)} (x-y).
\label{36}\ee
The scalar quark condensate can be extracted from small distance propagator behavior, $S(y)\equiv S(0,y)$ at small $y$, where $y$ may be purely temporal without loss of generality.
The free Green's function is
\be
S_0(y)= (m - \slashed\partial) \frac{m_q K_1(m_q |y|)}{4 \pi^2 |y|},
\label{37}\ee
where $K_1$ is the Macdonald function. For a massless fermion
\be
S_0(y) = \frac{\slashed y}{2 \pi^2 y^4},\quad m_q=0.
\label{38}\ee
According to the OPE for the quark propagator,
    the quark condensate is defined by the first order term in $\sigma$ $S_1(y)$ of the Green's function decomposition. 
\(S_1\) easily follows from \eqref{36} in the massless limit
\be
S_1(y)= \frac{\sigma}{8 \pi y},\quad m_q=0,
\label{39}\ee
while in the general case \cite{29}
\be
S_1(y)= \frac{\sigma m_q y K_1(m_q y)}{8 \pi y} + O( m_q y)
\label{40}\ee

At this point we need to find the effective distance $y$ that enters \eqref{37}-\eqref{40}. 
We extract the effective distance from the large-distance $S(y)$ asymptotics in the confining vacuum,
    which is $S(y)= \sum_n c_n \exp(-M_n |y|)$.
We estimate $y_{eff}$ as $\approx M_0^{-1}$
($M_0$ is the $q-\bar q$ mass without spin and perturbative interaction). 
For $m_q \ll M_0$, $M_0 \approx 1 $ GeV, hence $y_{eff}= 0.2 $ fm. 

Finally, we estimate the quark condensate
\be
\Delta_q= N_c \braket{S_1(y_{eff})}_{vac}= \frac{N_c \sigma M_0}{8 \pi}\frac{m_q}{M_0} K_1(m_q/M_0).\label{41}
\ee

In the massless quark limit the long-distance suppression is proportional to the string tension root $M_0\sim \sqrt{\sigma}$, 
    thus $\Delta\sim \sigma^{3/2}$, i.e.~the quark condensate is defined solely by the confinement.
$m_q=0$ and $ M_0= 1$ GeV yields $\Delta_l= (278 \text{ MeV})^3$, 
    which is close to our estimates in the previous sections, and to the lattice data.

For nonzero quark mass, the last factor in \ref{41} provides a small cut-off.
For example, for the strange quark $m_s= 100 $ MeV (at 2 GeV energy scale) and the same $M_0=1$ GeV, we obtained $\Delta_s=(276.6\text{ MeV})^3$ or $\Delta_s/\Delta_l= 0.985$. 
The estimate is near the lattice data \cite{30}, 
    which also shows a weak dependence on the strange quark mass value.
In the opposite limit of large quark mass $M_0\sim m_q$,
    the condensate is non-chiral $\Delta_{heavy}\sim \sigma m_q$.

\section{Conclusions and discussion}\label{SectConc}

We proposed a mechanism for the quark condensation in QCD, 
    and shown the defining role of the non-perturbative confining interaction for the effect.
The problem was addressed with several different approaches 
    that share the principal mechanism behind the effect,
        and essentially the same results were obtained.
Though our estimates account only for the leading non-perturbative contribution 
    (and we used any opportunity to resort to approximate calculations),
        the numerical results are close to the lattice data 
             --- all the values are within 10\% margin.

In section \ref{SectConf} we shown 
    that the quark propagator $S(x,y)$ definition in the confining vacuum involves the effective mass operator $M(x,y)$ \eqref{2}, \eqref{3},
        which dynamically brakes the chiral symmetry.
The effective mass enters as the basic factor in the expressions for $\Delta_i$ for any current quark mass.
         
In the next three sections, expressions \eqref{29}, \eqref{29'} for the quark condensate were obtained.
The confining string tension $\sigma$ enters the expressions as a factor of the spectral sum over PS non-chiral meson states.
The numerical results for the light flavors were obtained with a truncation of the spectral sum, 
    as well as with a well-defined regularization procedure (which, however, involves model assumptions and further approximations).

Of special importance is the previous section. 
The non-perturbative OPE analysis of the quark condensate provides automatic renormalization.
The resulting expression \eqref{41} is also applicable for nonzero quark mass. 
The condensate is chiral for light and strange flavors \(\Delta_{l,s}\sim \sigma^{3/2}\), and is proportional to the quark mass for a heavy flavor \(\Delta_h\sim\sigma m_h\).
The numerical values for light and strange quark condensates are very close,
    since the energy scale \(\sim 1\) GeV is set by a quark-antiquark system mass in the absence of spin-spin and perturbative corrections.

  \end{document}